\documentclass[prb, twocolumn, aps, superscriptaddress]{revtex4}

\newcommand{\bfk}{\mathbf k}
\newcommand{\bfkprime}{\mathbf k^{\prime}}
\newcommand{\bfq}{\mathbf q}

\newcommand{\ug}{UGe$_{2}$}

\usepackage{bm,epsf,graphicx}
\usepackage{hhline,subfigure,amsmath,verbatim,moreverb}

\expandafter\ifx\csname package@font\endcsname\relax\else
 \expandafter\expandafter
 \expandafter\usepackage 
 \expandafter\expandafter
 \expandafter{\csname package@font\endcsname}%
\fi

\begin{document}                % INITIALIZE - DONT CHANGE

% Title

\title{Ferromagnetic superconductivity driven by changing Fermi surface 
topology
}
\author{K. G. Sandeman, G. G. Lonzarich}
\affiliation{Low Temperature Physics Group, Cavendish Laboratory,
Madingley Road, Cambridge, CB3 0HE, United Kingdom.}
\author{A. J. Schofield}
\affiliation{School of Physics and Astronomy, University of
Birmingham, Edgbaston, Birmingham B15 2TT, United Kingdom.}
%\date{\today}
\begin{abstract}
We introduce a simple but powerful zero temperature Stoner model to
explain the unusual phase diagram of the ferromagnetic superconductor,
UGe$_{2}$.  Triplet superconductivity is driven in the ferromagnetic phase 
by tuning the majority spin Fermi level through one of two peaks in the
paramagnetic density of states (DOS). Each peak is associated with a
metamagnetic jump in magnetisation.  The twin peak DOS may be derived from
a tight-binding, quasi-one-dimensional bandstructure, inspired by previous
bandstructure calculations. 
\end{abstract}
\pacs{}
\maketitle

%%%%%%%%%%%%%%%%%%%%%%%%%%%%%%%%%%%%%%%%%%%%%%% 
%%%%%%%%%%% INTRODUCTION %%%%%%%%%%%%%%%%%%%%
%%%%%%%%%%%%%%%%%%%%%%%%%%%%%%%%%%%%%%%%%%%%%%% 
\section{Introduction}
\label{Intro}
The predominance of spin singlet superconductors over their triplet
counterparts has, in part, lead to the belief that superconductivity and
magnetism are mutually exclusive---the upper critical field of a singlet 
superconductor is bounded by the Pauli paramagnetic limit.  One 
might expect that a ferromagnet would be the natural stage for
spin triplet superconductivity, where we could overcome Pauli limiting 
effects.  But, until recently, there were no examples of
`ferromagnetic superconductivity' (FMSC)---the coexistence of itinerant
ferromagnetism (FM) and superconductivity (SC) in a single bulk 
phase~\cite{flouquet_2002a}.  This situation has
changed with the observation of FMSC in {\ug}~\cite{saxena_2000a},
URhGe~\cite{aoki_2001a} and ZrZn$_{2}$~\cite{pfleiderer_2001a}.  The
behaviour of these materials is an example of a more general phenomenon;
the observation of a novel state on the border of magnetism at low
temperatures.  By supressing magnetic order by some control parameter, be
it electron/hole density~\cite{lohneysen_2000a} or external
pressure~\cite{mathur_1998a} physicists are now able to access regimes
where magnetic fluctuations become quantum critical.

We choose here to concentrate on the case of {\ug}, because whilst SC is
only measurable in the ferromagnetic state (in common with URhGe and
ZrZn$_{2}$), {\ug} seems to posess particularly low electronic
dimensionality, uniaxial magnetisation, and revealing features in the
temperature-pressure phase diagram which we now review.
%%%%%%%%%%%%%%%%%%%%%%%%%%%%%%%%%%%%%%%%%%%%%%% 
%%%%%%%% UGE2 PHASE DIAG FIGURE %%%%%%%%%%%%%% 
%%%%%%%%%%%%%%%%%%%%%%%%%%%%%%%%%%%%%%%%%%%%%%%
\begin{figure} 
\includegraphics[width=0.9\columnwidth]{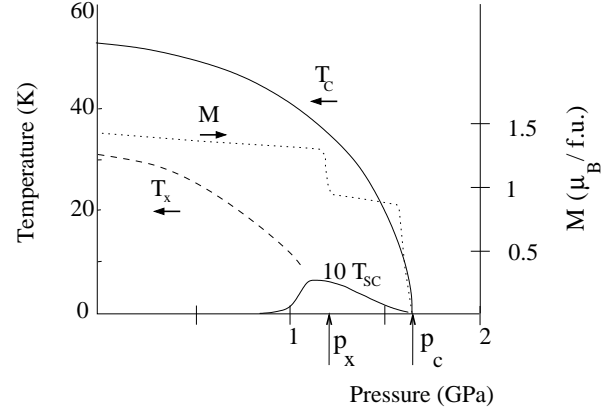}
\caption[Measured phase diagram of UGe$_{2}$ and experimental signatures
of $T_{x}$.]{The temperature-pressure phase diagram
~\cite{saxena_2000a,huxley_2001a,tateiwa_2001a} and low temperature easy 
axis magnetic moment, $M$ (after Pfleiderer and 
Huxley~\cite{pfleiderer_2001b}) of \ug. $T_{SC}$ is the
superconducting transition temperature, scaled by a factor of 10 for
clarity.  $T_C$ denotes the Curie temperature.  $T_x$ is a feature in the
ferromagnetic state seen in various measurements, described in the 
text.  $M$ shows two transitions; one at $p_x$, corresponding to where the 
$T_{x}$ line meets the pressure axis, and the other at the critical Curie 
pressure, $p_{c}$.}
\label{UGe2phaseFig} 
\end{figure}
%%%%%%%%%%%%%%%%%%%%%%%%%%%%%%%%%%%%%%%%%%%%%%% 
%%%%%%%%%%% PHASE DIAG SECTION %%%%%%%%%%%%%%%%
%%%%%%%%%%%%%%%%%%%%%%%%%%%%%%%%%%%%%%%%%%%%%%% 

In Fig. \ref{UGe2phaseFig} we show the temperature-pressure phase diagram
for {\ug}, with the Curie temperature $T_{C}$ (supressed to zero at
pressure $p_{c}$) and superconducting transition temperature $T_{SC}$ 
indicated~\cite{saxena_2000a,huxley_2001a,tateiwa_2001a}.  Another
feature, $T_{x}$ is also shown.  This $T_x$ shows up in measurements of
lattice expansion~\cite{oomi_1993a}, as a jump in the low temperature
$T^2$ component of the resistivity~\cite{saxena_2000a}, as a small
enhancement in specific heat~\cite{tateiwa_2001a}, as a kink in
resistivity~\cite{tateiwa_2001a} and as a change in the character of the
Fermi surface as measured in de Haas van Alphen
experiments~\cite{terashima_2001a}.  Most importantly for this work, $T_x$
also appears as a slight jump in magnetisation~\cite{tateiwa_2001a} which
is sharpened at lower temperatures such that the low temperature moment
has a step at pressure $p_x$ in addition to the step at the quantum phase
transition pressure, $p_c$ (see Fig. \ref{UGe2phaseFig}).   Furthermore, 
we note the close proximity of the $T_{x}(p)$ line to the peak in 
$T_{SC}$.

Most theories which describe ferromagnets close to a quantum phase
transition have predicted that the superconducting transition temperature,
$T_{SC}$ should be at least as high in the paramagnetic state as it is in
the ferromagnetic state.  These theories have considered a electronically
three-dimensional ferromagnet, either magnetically
isotropic~\cite{fay_1980a} or uniaxial~\cite{roussev_2001a}.  Where
theoretical models have predicted an enhancement of $T_{SC}$ in the
ferromagnetic regime, their basis seems unjustified in the case of {\ug}.  
Kirkpatrick and coworkers~\cite{kirkpatrick_2001a} have predicted an
enhancement of the superconducting $T_{SC}$ due to the coupling of magnons
to the longitudinal magnetic susceptibility.  However, the ferromagnetic
state of {\ug} is so magnetically anisotropic that the presence of
transverse magnons seems an unlikely primary explanation for the
enhancement of $T_{SC}$---at 4.2K and an external magnetic field of 4T,
the magnetisation along the easy axis is about 20 to 30 times that along
either of the other crystallographic axes~\cite{onuki_1992a}. 
Other authors have drawn their inspiration from bandstructure 
calculations.  Bandstructure 
analyses~\cite{shick_2001a,yamagami_1993a,tejima_2001a} of {\ug} seem to 
indicate that lowering temperature sparks the evolution of a 
quasi-two-dimensional majority carrier Fermi surface sheet below $T_C$. 
 Furthermore there is the possibility that large sections of the
quasi-two-dimensional Fermi surface may be parallel, making it almost
one-dimensional. Until now, this low-dimensional magnetism has pushed
authors in the direction of postulating the existence of a charge- or
spin-density-wave state (CDW or SDW) below
$T_{x}$~\cite{saxena_2000a,huxley_2001a}, sometimes by analogy with the
$\alpha$ phase of uranium.  Watanabe and Miyake~\cite{watanabe_2001a} have
postulated that the interplay of CD or SD fluctuations at high wavevector
will couple to the magnetisation, $M$ in such a way as to enhance it at
some critical value, $M_{x}$\footnote{By coupling
high wavevector fluctuations to the ${\bfq}=0$ mode, Watanabe and Miyake's
model is reminiscent of the spin-bag mechanism of nodeless
superconductivity proposed by Schrieffer et al. in the context of the
cuprates~\cite{schrieffer_1989a}.}. However, spin density fluctuations 
have yet to be observed in neutron experiments on \ug.

We turn to the low dimensional bandstructure for a different effect.  The
key idea will be that in a ferromagnet, somewhat uniquely, the
magnetisation acts as a tuning parameter which can subtly change the
topology of the {\it anisotropic} Fermi surfaces of different spin
species.  By contrast, in a rigid band picture of a paramagnetic metal,
the Fermi surface is fixed.  The added topological possibilities for a
ferromagnet should be viewed as a useful tool---and as a reason for
observing the enhancement of features, such as $T_{SC}$, within the
ferromagnetic phase. This paper is planned as follows: firstly, we show
that an electronic density of states (DOS) which has two peaks can
reproduce the two steps in the observed low temperature magnetisation.  
We then show that the necessary form of DOS arises naturally from a low
(quasi-one-) dimensional bandstructure and that the magnetisation 
resulting from this bandstructure has a jump in the ferromagnetic state 
which is coincident with the maximum in a superconducting instability, 
mediated by spin fluctuations.

%%%%%%%%%%%%%%%%%%%%%%%%%%%%%%%%%%%%%%%%%%%%%%% 
%%%%%%%%%%%%%%%  MODEL			%%%%%%%
%%%%%%%%%%%%%%%%%%%%%%%%%%%%%%%%%%%%%%%%%%%%%%% 
\section{Model}
\label{Model}
The simplest Stoner theory of magnetism is long-known to be inadequate in
describing the temperature dependence of the magnetisation even of
magnetically isotropic, electronically 3-dimensional ferromagnets, where
spin fluctuations have to be taken into account.  There, a
fluctuation-averaged equation of state method, in the spirit of Lonzarich
and Taillefer~\cite{lonzarich_1985a} or Yamada~\cite{yamada_1993a} would
be a improved model of magnetism at finite temperature, where we expect
the temperature dependence of $M$ to arise from the fluctuation response,
rather than just the Fermi functions included in Stoner theory.  The 
pronounced magnetic anisotropy of {\ug} might ordinarily simplify matters, 
as the consequent absence of transverse spin modes will make a Stoner
approach more valid, especially at low temperatures.  However the reduced
electronic dimensionality of {\ug} will probably heighten the importance 
of spin fluctuations at finite temperatures.

We therefore circumvent finite temperature concerns by employing a {\it
zero} temperature Stoner theory, with the first aim being to reproduce the
step in $M(p)$ at $p_{x}$. We consider the action of pressure to be akin 
to that of varying the
exchange energy, $I$ in a Stoner model of the one-electron energy of
separated majority (say, $\uparrow$) and minority (say, $\downarrow$) spin 
sheets: 
%\begin{equation}
$E_{k\sigma}=\epsilon_{k}\pm I M \,\,\,\,\, (-\uparrow, +\downarrow).$
%\label{}
%\end{equation}  
In this description, the occupation of each spin sheet $\sigma$ is
%\begin{equation}
$n_{\sigma}=\int_{\epsilon_{b}}^{\mu_{\sigma}}\rho(\epsilon)d\epsilon$,  
%\, , \, \,
where
$M={1 \over 2}(n_{\uparrow}-n_{\downarrow})$.
%\end{equation}
%
Here $\epsilon_{b}$ is the bottom of the band, and spin $\sigma$ 
occupies energy
states up to $\mu_{\sigma}$.  The DOS is given
by $\rho(\epsilon)$.  We consider the total number of spins,
$n_{\uparrow}+n_{\downarrow}=N$ to be fixed.  The chemical 
potential, $\mu_{\sigma}$ of each spin sheet is therefore completely
determined by the particular
magnetisation and chosen electron number.  The total energy 
density of the electron system is
\begin{eqnarray}
F[M]=\int_{\epsilon_{b}}^{\mu_{\uparrow}}\epsilon\rho(\epsilon)d\epsilon+
\int_{\epsilon_{b}}^{\mu_{\downarrow}}\epsilon\rho(\epsilon)d\epsilon 
\nonumber \\+
I\bigl({N^2 \over 4} -M^2 \bigr) -g\mu_{B}H M,
\label{FreeEnergy}
\end{eqnarray}
where we have included a term for the presence of an external magnetic
field, $H$.  

\begin{figure}
\includegraphics[width=\columnwidth]{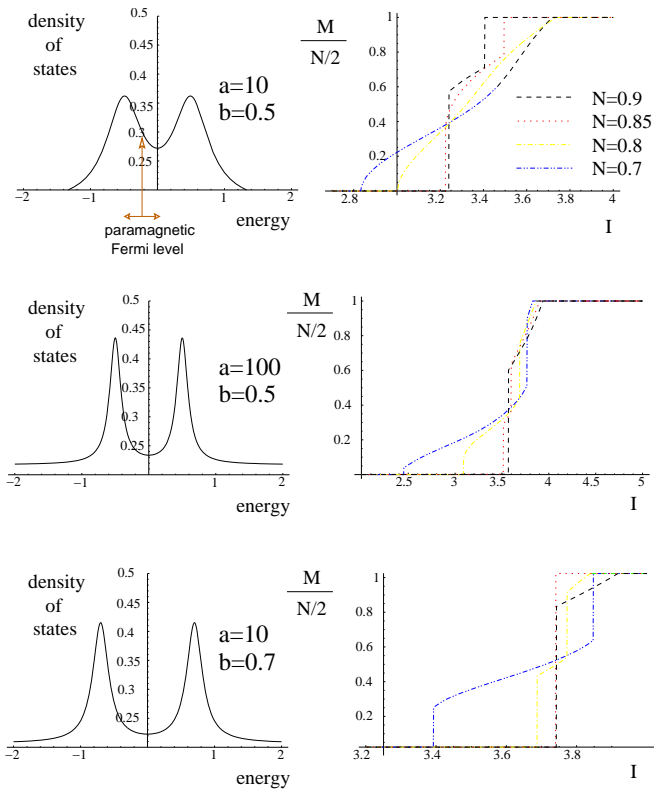}
\caption[How a DOS with two peaks can give two transitions in the
magnetisation of a ferromagnet within Stoner theory]
{Using the twin-peak density of states given by Eqs. \ref{lor1} and
\ref{lor2}, it is possible to obtain two transitions in the
magnetisation of a Stoner ferromagnet, where the exchange parameter $I$
is the varied quantity.  We plot both the normalised density of states
and the resultant $M(I)$ curve.  If the two peaks in the DOS are too far
apart, a saturated magnetic state is favoured.  If the two peaks are
too smooth, the transitions in $M(I)$ are weakened.  The
bandstructure parameters used in each case are indicated, together with  
the range of number of spins, $N$, used in the calculations of $M(I)$.}
\label{CombiLorFig}
\end{figure}

Most phenomenological expansions of this energy density have included
terms even in $M$, up to order $M^6$.  This can give
one first order transition in $M(I)$.  However, we are looking for an
additional transition, corresponding to $p_{x}$, and believed to be first
order~\cite{pfleiderer_2001b}, although there is some controversy over 
this~\cite{terashima_2002a}.  To have the possibility of an additional 
{\it first} order transition, we need the next even term in
the Landau free energy expansion, thus obtaining an $M^8$
theory~\cite{shimizu_1982a}.    We now show that a DOS with two peaks 
generically brings about the $M^{8}$ term in $F[M]$ by allowing a 
scenario where the global magnetisation can change rapidly, twice.
We begin by taking a one-band DOS comprised of two Lorenztians, 
normalised such that the maximum number of electrons of each spin in the 
band is 1:  

\begin{equation}
\rho(\epsilon)={\rho_{0}(\epsilon)\over\int_{\epsilon_{B}}^{\epsilon_{T}}{\rho_{0}(\epsilon)d\epsilon}}
\label{lor1} 
\end{equation}
where
\begin{equation}
\rho_{0}(\epsilon)=1+{1\over a(\epsilon-b)^2 +1}+{1\over
a(\epsilon+b)^2 +1}\, ,
\label{lor2}
\end{equation}
where we can vary $a$ to adjust the sharpness of the two humps and $b$ to
adjust their position.  The humps are symmetric with respect to the zero 
energy and are centred on $\pm b$.  We minimise $F[M]$, given by
Eq. \ref{FreeEnergy} with respect to magnetisation, $M$ to obtain the
variation of $M$ with respect to the Stoner exchange, $I$.
In Figure \ref{CombiLorFig} we show plots of $M(I)$ for three sets of
parameters $a$ and $b$, in each case setting the bottom and top,
$\epsilon_{B}$ and $\epsilon_{T}$ of the band to be $-2$ and $+2$ 
respectively.
%%%%%%%%%%%%%%%%%%%%%%%%%%%%%%%%%%%%%%%%%%%%%%%
%%%%%%%% UGE2 LOR BANDS FIGURE   %%%%%%%%%%%%%%
%%%%%%%%%%%%%%%%%%%%%%%%%%%%%%%%%%%%%%%%%%%%%%%
For each set of parameters, we show the effect of varying the paramagnetic
Fermi level, that is to say, the number of spins, $N$ in the system, in 
the neighbourhood of half-filling.  This places the paramagnetic Fermi 
level, $\mu_{EF}$ between the two DOS peaks.  On increasing the
exchange term, $I$, the spin sea is polarised and the majority Fermi level
moves up from that of the paramagnetic state and the minority level moves
down.  The onset of ferromagnetism is governed by the usual Stoner 
criterion, $I\rho(\mu_{EF})=1$ in the case of a second order onset, or by 
the finite-magnetisation Stoner criterion \cite{fazekas_1999a} for a first order onset:
\begin{equation}
{2 I M \over \mu_{\uparrow}-\mu_{\downarrow}}=1.
\label{FiniteStoner}
\end{equation}
The finite magnetisation criterion holds for all $M$ once the system is  
polarised, irrespective of the order of the onset.
Each Fermi level feels the effect of the density of states peaks at
different $I$, as long as the paramagnetic Fermi level is
off-centre with respect to the two DOS peaks. The main result is that, by
making the density
of states peaks sufficiently sharp and close together, {\it we can obtain
two magnetic transitions, one from the paramagnetic state and another
within the ferromagnetic state.} If the DOS peaks are too far apart, a
saturated magnetic state is favoured; if the peaks are too smooth, the
transitions in $M(I)$ are weakened and if
the starting paramagnetic level is too close to the half-way point between
the two peaks ($N$ too close to 1) then only one transition is observed.
Therefore, to observe two transitions, the paramagnetic filling level
should be between the DOS peaks, but off-centre with respect to them.

%%%%%%%%%%%%%%%%%%%%%%%%%%%%%%%%%%%%%%%%
%%%%%%% Bandstructure subsection %%%%%%%
%%%%%%%%%%%%%%%%%%%%%%%%%%%%%%%%%%%%%%%%
\subsection{Bandstructure phenomenology}
Having shown that a double-humped DOS is perhaps key to understanding the
magnetic properties of {\ug} at low temperatures, we now set about working
towards a tight-binding picture of the bandstructure of this compound
which can reproduce both its magnetic and superconducting properties.
There have been two major efforts towards bandstructure calculations of
{\ug}.  The first,
originally on a crystal structure with an incorrect
space group was performed by Yamagami et al.~\cite{yamagami_1993a} and has
since been revised~\cite{tejima_2001a}. This produced spin-separated Fermi
sheets in the ferromagnetic state, centred on the $\Gamma$ point, with the
majority sheet strongly nested. The second approach, already mentioned,
has been an LDA $+U$ method, by Shick and Pickett~\cite{shick_2001a},
using $U$ as a fitting parameter for the zero temperature, ambient   
pressure magnetisation.
Again, strongly nested Fermi sheets were obtained, but they were of mixed
spin type and the sheet considered by those authors as being most
important for superconductivity is centred on the crystallographic $M$
point in their model.  Furthermore, there was
initial disagreement on the direction of the nesting vector ${\mathbf Q}$
relative to the easy axis of magnetisation ${\mathbf a}$ --- both are
`in-plane', but Yamagami considered ${\mathbf Q}\perp{\mathbf a}$ whilst
Shick found ${\mathbf Q}\parallel{\mathbf a}$.  In summary, we take the
following minimal, but key ingredients in our model:
\begin{itemize}
\item quasi-one-dimensional bandstructure (crucial to what follows)
\item strong uniaxial magnetic anisotropy and
\item spin-split Fermi sheets (even in the LDA $+U$
work~\cite{shick_2001a}, there is a large exchange splitting, of the
order of 1 $eV$).
\end{itemize}   

In our model, we utilise the interaction potential for spin fluctuation   
mediated pairing in the ferromagnetic state, as derived by Fay and
Appel~\cite{fay_1980a}.  We 
also
follow their sign convention, namely that an attractive potential
between like spins is positive.  The interaction potential is heavily
dependent on the Lindhard response, $\chi^{(0)}_{\sigma\sigma}(\bfq)$
for the bandstructure under consideration:
\begin{equation}
V_{\sigma\sigma}({\mathbf q})={I^{2} \chi^{(0)}_{-\sigma
-\sigma}({\mathbf q})
\over
1-I^{2}\chi^{(0)}_{\sigma\sigma}({\mathbf q})\chi^{(0)}_{-\sigma
-\sigma}({\mathbf q})},
\label{FayAppelV}
\end{equation}
where $I$ is again the repulsive Hubbard-type contact interaction acting
between opposite spins and $\chi^{(0)}_{\sigma\sigma}(\bfq)$ is given by
\begin{equation}
\chi^{(0)}_{\sigma\sigma}({\mathbf
q})=\sum_{\bfk}{f^{(0)}_{\sigma}(\bfk)-f^{(0)}_{\sigma}(\bfk+{\mathbf
q})\over
\epsilon (\bfk+{\mathbf q})-\epsilon (\bfk)}\, ;
\label{Lindhard}
\end{equation}
$f^{(0)}_{\sigma}(\bfk)$ being the Fermi occupation function for spin
$\sigma$ which is at chemical potential $\mu_{\sigma}$ ($\mu_{\sigma}$ is
defined previously).  According to our chosen interaction mechanism, Eq.
\ref{FayAppelV}, a
large $\chi^{(0)}_{\sigma\sigma}(\bfq)$ might naively be expected to
lead to an
enhanced $T_{SC}$.
However, the subtlety here is that the interaction potential in the
ferromagnetic state mixes longitudinal susceptibilities of majority and
minority spin sheets so the effect is not so clear-cut.  We recall that in
the free electron model, only in electronic dimensions less than two,
there can be a peak in $\chi^{(0)}_{\sigma\sigma}(\bfq)$ at non-zero
$\bfq$~\cite{gruner_1994a}.  In that case, we might expect that when one
sheet of the Fermi surface is at optimal nesting, then perhaps
$V_{\sigma\sigma}({\mathbf
q})$ will be highest.  The problem with this argument
is that a $V_{\sigma\sigma}(\bfq)$
dominated by high-${\mathbf Q}$ modes would normally lead to very weak
triplet pairing and in tight-binding approaches, finite $\bfq$ peaks in
$\chi^{(0)}_{\sigma\sigma}(\bfq)$ are possible in two and three
dimensions.

\begin{figure}
\includegraphics[width=\columnwidth]{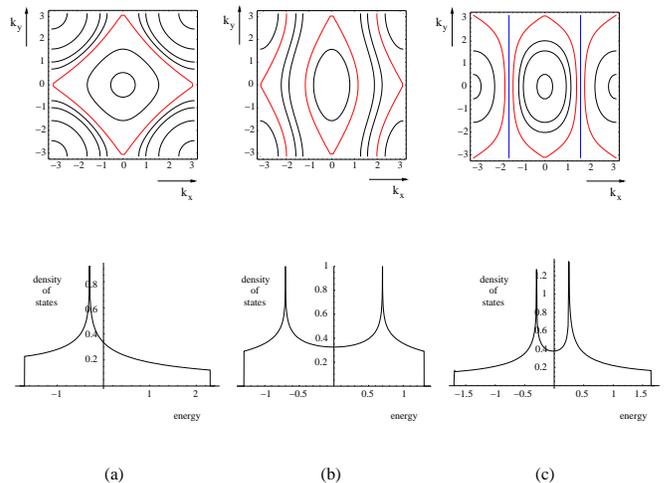}
\caption[Contour plots and densities of states of two-dimensional and
quasi-one-dimensional tight-binding bandstructures.]
{Two dimensional and quasi-one-dimensional tight-binding
bandstructures: contour plots (top row of figures) with the van Hove
contours shown in red and associated densities of states (bottom row).
In (a) the two-dimensional structure given by Eq. \ref{2Dstruct} with
$\alpha_{x}=\alpha_{y}=1$ and $\beta=-0.3$ is shown.  This is a
structure
most often studied in
connection with the cuprates.  In  (b) we show the quasi-one-dimensional
structure of Eq. \ref{Org1D} with $\alpha_{x}=1,\alpha_{y}=0.3$.  In (c),
higher harmonics in the principal direction are included (Eq.
\ref{ugmyband}) with $\alpha_{x}=1$, $\beta=0.7$, $\delta=0.03$ and
$\gamma=-0.03$.  We see that in (a) there is one (degenerate) van Hove 
contour, but on going into quasi-one-dimensions,
we lift this degeneracy and find two van Hove contours and hence two
maxima in the density of states.  Furthermore, the
choice of bandstructure in (c) has an optimal nesting scenario (shown
in blue) which is
separate from both of the van Hove contours, due to the elimination of the
$\cos k_{y}$ term in the bandstructure.}
\label{vanNestingFig}
\end{figure}

The approach we take is to look for density of states-driven
superconductivity, where the large density of states giving rise to the
magnetisation step at $M_x$ is also able to enhance superconductivity in
the ferromagnetic state.  In tight-binding theory in two dimensions and  
lower, the peak in the density of states comes from the presence of a van
Hove singularity.   The simplest, familiar tight-binding bandstructure for
the cuprates is of the form
\begin{equation}
\epsilon({\bfk})=-\alpha_{x}\cos k_{x}-\beta\cos k_{x}\cos
k_{y}-\alpha_{y}\cos
k_{y}
\label{2Dstruct}
\end{equation}
with $\alpha_{x}=\alpha_{y}=1$ and $\beta$ less than 1.  This
corresponds to equal nearest
neighbour hopping in each of the two dimensions, and includes the
next-nearest neighbour term, $\cos k_{x}\cos k_{y}$.  This structure, as
shown in Figure \ref{vanNestingFig}(a), has one van Hove singularity and
is therefore of little use in studying the magnetic properties of
{\ug}---as we have already found, we require {\it two} density of states
peaks.  However, as we will now show, if we now reduce the amplitude of
the $\cos k_y$ term, we  cross over into quasi-one-dimensions and begin to
lift the degeneracy of the van Hove contours, giving two of the them,
with adjustable separation.
%%%%%%%%%%%%%%%%%%%%%%%%%%%%%%%%%%%%%%%%%%%%%%%
%%%%%%%% van Hove and Nesting FIG $$$$%%%%%%%%%
%%%%%%%%%%%%%%%%%%%%%%%%%%%%%%%%%%%%%%%%%%%%%%%

The most common quasi-1D bandstructure in the literature is probably
\begin{equation} \epsilon({\bfk})=-\alpha_{x}\cos k_{x}-\alpha_{y}\cos
k_{y} \label{Org1D}\end{equation} which is used to describe some organic
superconductors.  Here, $\alpha_{x}$ is 1 and $\alpha_{y}$ is less than 1
and the model corresponds to nearest neighbour hopping in each of the two
dimensions.  This bandstructure has two van Hove contours (see Figure
\ref{vanNestingFig}(b) and hence two density of states peaks.  Thus, on
tuning our magnetisation, we can observe in both spin types the transition
through a van Hove singularity in the density of states at the Fermi
level, provided, as stated before, that our paramagnetic Fermi level is 
between the two peaks in the density of states.  The problem arises that
when the peaks are close enough together to cause an effect in the
magnetisation curves, the variation of the DOS in between them is not   
very rapid.  This leads to a weakening of the transitions in $M(I)$ as 
found with a slowly varying Lorentzian DOS in the last section.

However, if we assume a quasi-one-dimensional dispersion of the form

\begin{equation} \epsilon({\bfk})=-\alpha_{x}\cos k_{x}-\beta\cos
k_{x}\cos k_{y}-\gamma\cos{2 k_{x}}-\delta\cos{3 k_{x}}
\label{ugmyband}
\end{equation}
with $\alpha_{x}=1$ and
$\beta$,$\gamma$, and $\delta$ all less than unit magnitude, {\it we will
be
able to explore the possibility of two, first order jumps in $M(I)$}.

The higher harmonics in the principal direction $k_{x}$ and the
next-nearest term, $\cos k_{x}\cos k_{y}$ which appear in Eq.
\ref{ugmyband} are not unreasonable in a system which we know to be
strongly one-dimensional in character.  The nearest neighbour term in
$k_{y}$ has been reduced to zero.  As shown in Figure
\ref{vanNestingFig}(c), this dispersion relation yields two van Hove
contours (and hence maxima in electronic density of states), and also goes
through a perfect nesting scenario.\footnote{It is worth noting that, on 
scanning through energy in {\it quasi}-one-dimensions, optimal nesting 
(which is also the density of
states minimum) does {\it not} coincide with the van Hove point (the
density of states maximum).  This is because optimal nesting in quasi-1D
results from a Fermi surface consisting of parallel lines, each parallel
to one axis of the Brillouin zone, as shown by the blue lines in Figure
\ref{vanNestingFig}(c).  By contrast, the optimal nesting Fermi surface of
a two-dimensional square lattice in a nearest-neighbour tight-binding
model is a diamond shape, with parallel surfaces at 45 degrees to the
Brillouin zone axes.  As shown in Figure \ref{vanNestingFig}(a), this 
Fermi surface also corresponds to the van Hove point and density of states
maximum.}  We note that our choice of bandstructure is based on an 
extrapolation from
the one point ($T=0$, $p=0$) of the phase diagram calculated in Refs.
\cite{shick_2001a,yamagami_1993a,tejima_2001a}, with the idea being that
there should be strong nesting present at full magnetisation in our model,
to match the bandstructure calculations.  Indeed, the high energy contours
of this dispersion are fairly well nested (Figure \ref{vanNestingFig}(c)).
To achieve a magnetically saturated state which does not completely fill  
the band, we require less than half-filling of the band in the
paramagnetic state, a requirement which is in line with the demands of the
$M(I)$ profile from Section \ref{Model}.  Our choice of bandstructure will
ultimately help to link
$p_x$, the maximum in $T_{SC}$ and the mass enhancement observed in de
Haas van Alphen measurements.  First we concentrate on the magnetisation  
as a function of Stoner exchange, $I$.

%%%%%%%%%%%%%%%%%%%%%%%%%%%%%%%%%%%%%%%%%%%%%%%%%%%%%%%%%%%%%%%%%%%%%%
%%%%%%%%%%%%%%%%%%%%%%%%%%%%%%%%%%%%%%%%%%%%%%%%%%%%%%%%%%%%%%%%%%%%%%
%%      RESULTS Getting a first order jump in M with I              %%
%%%%%%%%%%%%%%%%%%%%%%%%%%%%%%%%%%%%%%%%%%%%%%%%%%%%%%%%%%%%%%%%%%%%%%
%%%%%%%%%%%%%%%%%%%%%%%%%%%%%%%%%%%%%%%%%%%%%%%%%%%%%%%%%%%%%%%%%%%%%%
\section{Results}
\subsection{Two jumps in magnetisation from a tight-binding bandstructure}  
\label{jumps}
\begin{figure}
\includegraphics[width=0.95\columnwidth]{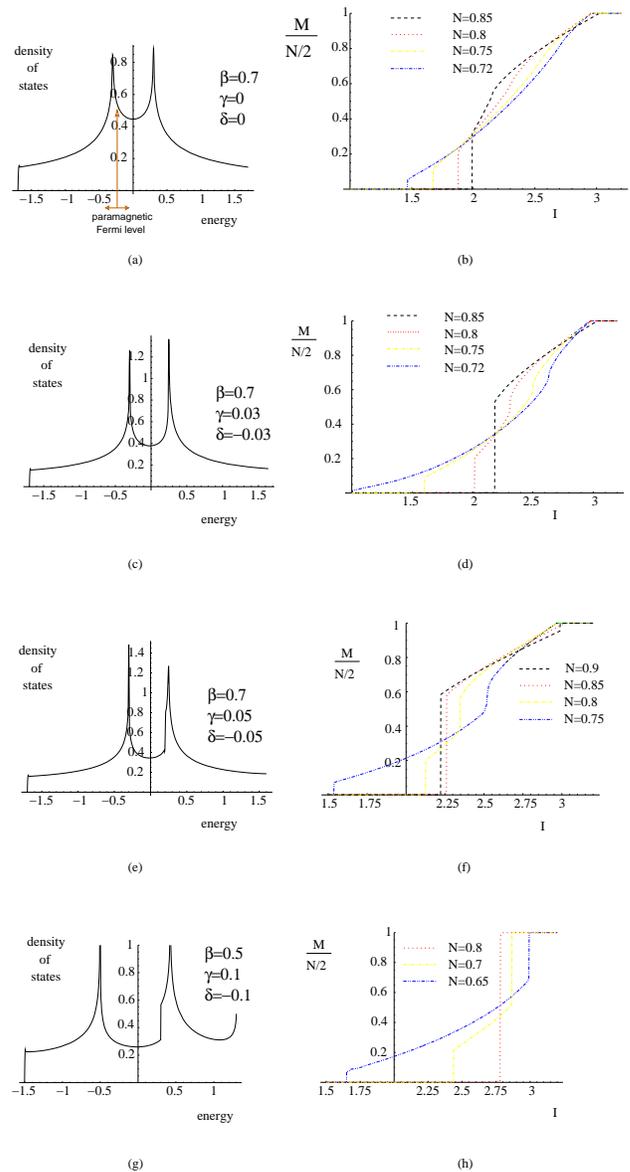}
\caption[Quasi-one-dimensional tight-binding densities of electronic 
states,
and resultant zero temperature M(I) curves within Stoner theory.]
{A set of phenomenological quasi-one-dimensional dispersions for
UGe$_2$,
where both the possibility of tuning the magnetisation through a van Hove 
singularity and a perfect nesting condition are present.
(a), (c), (e) and (g) show
the calculated density of states whilst (b), (d), (f) and (h) are the
calculated graphs of magnetisation as a function of $I$, the Stoner
exchange parameter for 
various levels of band-filling, below half-filling.  The tight-binding
parameters for our dispersion, $\epsilon({\bfk})=-\cos k_{x}(1+\beta\cos  
k_{y})-\gamma\cos{2 k_{x}}-\delta\cos{3 k_{x}}$ are indicated.  As in 
Fig. \ref{CombiLorFig}, two magnetic transitions are often visible.}
\label{UGe2MyBandFig}
\end{figure}

Taking the quasi-one-dimensional bandstructure given by Eq.
\ref{ugmyband} and varying
parameters $\beta$, $\gamma$, and $\delta$, we obtain the variation of    
zero temperature magnetisation with exchange
interaction $I$ as we did for Lorentzian densities of 
states.\footnote{Setting $\alpha_{x}$ to 1 in Eq. \ref{ugmyband} and
varying $I$ is equivalent to varying the quantity 
$I^{\prime}=I/\alpha_{x}$, as can be
seen from the form of the free energy in Eq. \ref{FreeEnergy}.  The effect
of increasing pressure is to increase, probably most strongly, the hopping
parameter $\alpha_{x}$.  This suppresses ferromagnetism, as it reduces
$I^{\prime}$.  We show plots of $M(I)$ since $I=I^{\prime}$ here.}
In Figure \ref{UGe2MyBandFig}, we present $M(I)$ for different total  
numbers of electrons, $N$ below half-filling in the case of four
different sets of tight-binding parameters. In each case, the paramagnetic
Fermi level sits in between the two peaks in the DOS. Initial polarisation
at $I_c$ (labelled by analogy with $p_c$ in experiment) is first 
order---this
is due to the
minority sheet approaching the lower van Hove peak as polarisation 
commences.
%%%%%%%%%%%%%%%%%%%%%%%%%%%%%%%%%%%%%%%%%%%%%%%
%%%%%%%% UGE2 MY BANDS FIGURE    %%%%%%%%%%%%%%
%%%%%%%%%%%%%%%%%%%%%%%%%%%%%%%%%%%%%%%%%%%%%%%

There are several features to be noted from the plots in Fig.
\ref{UGe2MyBandFig}.  Firstly, as in Fig. \ref{CombiLorFig}, if $N$ is
below
half-filling, a second magnetic transition can occur when the majority
spin sheet feels the effect of the upper peak in the DOS.  However, when N
is too close to half filling, or when the dip between the two density of
states peaks is too sharp, we get only one first order transition, which    
can even be straight to saturation magnetisation.
Once again, in order to observe the {\it two} transitions seen in \ug, one
should have a paramagnetic Fermi level which is off-centre with respect to
two DOS peaks.
Secondly, the upper transition (which we henceforth label $I_{x}$ by
analogy with $p_{x}$ from experiment), can either be to finite
magnetisation (and either first order or beyond the first order critical  
point) or can be first order to saturation magnetisation.  According to   
Eq. \ref{FayAppelV}, a transition to saturation magnetisation would   
naturally kill any magnetically-mediated superconductivity due to the lack
of one spin species.  Thus we turn our attention to an upper transition
not of saturating nature, and which could be first order, or close to
first order, as the latter would allow the Fermi surface sheet
configurations to be explored as the magnetisation is tuned by temperature
or pressure, allowing the presence of sharp peaks in the density of states   
to be observed in measurements such as heat capacity and resistivity.

The above considerations make the second example in Fig.
\ref{UGe2MyBandFig} a most useful candidate, as we have the option of the
upper transition being first order or close to first order, depending
on $N$, and the magnetisation does not saturate at the upper transition.
Henceforth, we take $\beta=0.7$, $\gamma=0.03$ and $\delta=-0.03$.

\begin{figure}
\includegraphics[width=0.9\columnwidth]{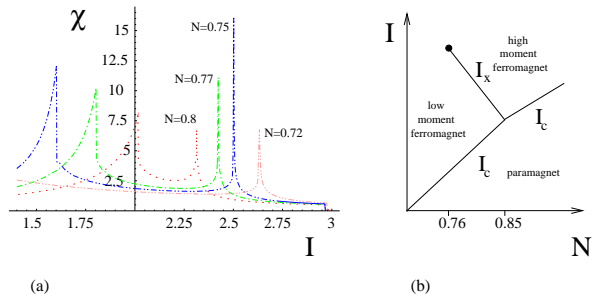}
\caption[Magnetic susceptibility of our model and progression towards
critical point associated with $I_x$.]{The phase diagram of our model in     
$N,I$ space shows a line of first order transitions at $I_{x}(N)$ which
end in a critical point.  The transitions at the Curie onset, $I_{c}$ are 
also
first order.  This can be seen in the behaviour of (a) the
longitudinal susceptibility and (b) the resulting (approximate) phase
diagram.  The tight-binding parameters used here are  $\beta=0.7$,   
$\gamma=0.03$ and $\delta=-0.03$, in line with what follows.}
\label{CombiSuscFig}
\end{figure}

Interestingly, the magnetic phase diagram in the $N,I$ plane seems
to contain a line of first order transitions corresponding to   
$I_{x}(N)$.  This line terminates at a critical point at a value of $N$ of
about 0.76.  For values of $N$
below this, the transition in $M(I)$ is smooth and can be described as
close to first order.  This is perhaps best seen in Figure
\ref{CombiSuscFig}(a), where we show the longitudinal susceptibility,
$\chi$
as a function of $I$ for $\beta=0.7$, $\gamma=0.03$ and $\delta=-0.03$,
taking different values of $N$.  We take the longitudinal susceptibility
from the Stoner forms:
\begin{center}
\begin{eqnarray}
\label{StonerChiPM}
\chi_{para}={\rho(\mu_{EF}) \over 1-I\rho(\mu_{EF})}\, ,\\
\label{StonerChiFM}
\chi_{ferro}={2\rho(\mu_{\uparrow})\rho(\mu_{\downarrow}) \over
\rho(\mu_{\uparrow})+\rho(\mu_{\downarrow})-2I
\rho(\mu_{\uparrow})\rho(\mu_{\downarrow})}
\label{StonerChi}
\end{eqnarray}
\end{center}
in the paramagnetic (Fermi level $\mu_{EF}$) and ferromagnetic states,
respectively.  As seen in Figure
\ref{CombiSuscFig}(a), the susceptibility at $I_x$ grows as we move closer 
to the critical point;  the resulting (approximate) phase diagram in 
$(N,I)$
space is shown in Figure \ref{CombiSuscFig}(b).
%%%%%%%%%%%%%%%%%%%%%%%%%%%%%%%%%%%%%%%%%%%%%%%
%%%%%%%% COMBI SUSCEPT  FIGURE   %%%%%%%%%%%%%%
%%%%%%%%%%%%%%%%%%%%%%%%%%%%%%%%%%%%%%%%%%%%%%%

\subsection{Magnetic field effects}
\begin{figure}
\includegraphics[width=\columnwidth]{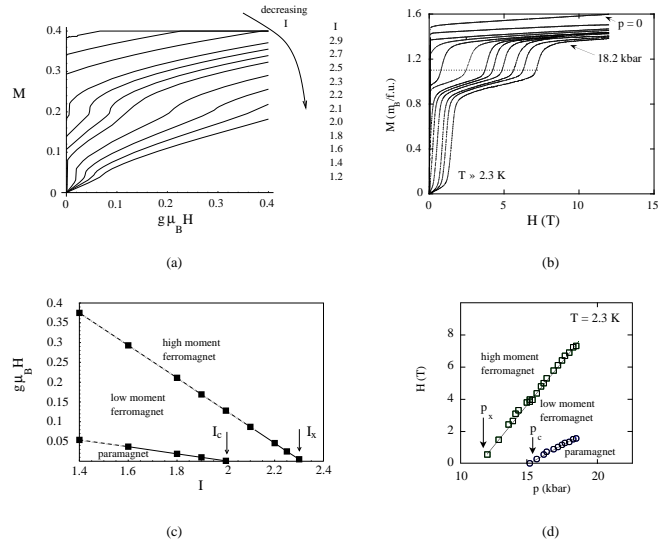}
\caption[Pushing the jump in $M(I)$ to lower $I$ by applying a magnetic
field.]{Our model predicts that both the magnetic jump at the
onset of ferromagnetism and that at finite
magnetisation can be pushed to lower values of exchange energy $I_c$ or   
$I_{x}$ respectively (corresponding to pressures higher than $p_c$ and 
$p_x$) by the application of a magnetic field, $H$.  The
tight-binding parameters used here are again $\beta=0.7$,
$\gamma=0.03$ and $\delta=-0.03$.  The number of spins, $N$ is 0.8 in both
figures (a) and (c) which show the calculated $M(I)$ curves and the
resulting phase diagram in $H,I$ space.
Recent experimental data, from Pfleiderer and Huxley
\cite{pfleiderer_2001b} is shown in (b) and (d).  This has a
form similar to that predicted in (a) and (c),
with the appearance of two transitions in $M$, which are tunable with 
pressure.}
\label{MagFieldEffectFig}
\end{figure}

It has also been found that the features associated with $T_{x}$ and $T_c$
can be recovered at pressures above $p_{x}$ and $p_c$ respectively by the
application of a magnetic field.  In our model, this metamagnetism 
arises as a direct consequence of
having spin-split Fermi surfaces, with the spin species being governed by
number conservation and the free energy of the form given in Eq.
\ref{FreeEnergy}.  As shown in Fig. \ref{MagFieldEffectFig}, turning on
the magnetic field, $H$, pushes both the magnetisation jump at $I_c$ (the
Curie transition) and at $I_x$ (within the ferromagnetic state) to lower
values of $I$, or equivalently, higher pressures.  Both the predicted
$M(I)$ curves and the resulting phase diagram in $H,I$ space bear a
striking resemblance to recent experimental data~\cite{pfleiderer_2001b}.
We should point out that the calculated $H,I$ phase diagram was obtained
by
looking for the maximum in the gradient of the magnetisation for each
$M(I)$ plot in field.  The first order nature of both transitions is
softened with decreasing $I$---corresponding to going to higher
pressures.  Experimentally, such softening is not observed at pressures
reached this far, as can be seen from Figure \ref{MagFieldEffectFig}(b).
We indicate where the magnetic transition in our model is no longer first
order by a dotted line on the $H,I$ phase diagram.

 %%%%%%%%%%%%%%%%%%%%%%%%%%%%%%%%%%%%%%%%%%%%%%%
%%%%%%%% MAGNETIC FIELD FIGURE   %%%%%%%%%%%%%%
%%%%%%%%%%%%%%%%%%%%%%%%%%%%%%%%%%%%%%%%%%%%%%%
%%%%%%%%%%%%%%%%%%%%%%%%%%%%%%%%%%%%%%%%%%%%%%%%%%%%%%%%%%%%%%%%%%%%%%
%%%%%%%%%%%%%%%%%%%%%%%%%%%%%%%%%%%%%%%%%%%%%%%%%%%%%%%%%%%%%%%%%%%%%%
%%              Superconductivity and phase diagram                 %%
%%%%%%%%%%%%%%%%%%%%%%%%%%%%%%%%%%%%%%%%%%%%%%%%%%%%%%%%%%%%%%%%%%%%%%   
%%%%%%%%%%%%%%%%%%%%%%%%%%%%%%%%%%%%%%%%%%%%%%%%%%%%%%%%%%%%%%%%%%%%%%
\subsection{Superconductivity and the zero-temperature phase diagram}  
From the heavy nature of the quasiparticles alone, we expect strong
coupling theory to be required to gain a true estimate of $T_{SC}$ in   
{\ug}.  However, we can gain useful information on the form of the
phase diagram from examining the zero temperature, weak-coupling   
properties of our model.  We note here that, without a reliable model of 
$M(T)$, it is impossible to construct the usual $T_{SC}$ equation in the
BCS theory in the ferromagnetic state, as here we have an example of a
temperature-dependent potential---the changing magnetisation alters the
spin-dependent interaction.

A reasonable estimate of the strength of triplet pairing can be gained 
from examining the relative magnitudes of the mass renormalisation
parameter $\lambda_{Z}$ and interaction parameter $\lambda_{\Delta}$,
defined as
\begin{center}
\begin{eqnarray}
\label{ldaz}
\lambda_{Z}={\int_{FS}\,\int_{FS'} \, d^{2}k\,d^{2}k^{\prime} \,
V_{\uparrow\uparrow}({\bfk}-{\bfkprime})
\over
\int_{FS} d^{2}k}\, ,\\
\label{ldad}
\lambda_{\Delta}={\int_{FS}\,\int_{FS'}\,d^{2}k\,d^{2}k^{\prime}
V_{\uparrow\uparrow}({\bfk}-{\bfkprime})  
\eta({\bfk})\eta({\bfkprime})\over
\int_{FS} d^{2}k \, \eta^2({\bfk})}\, ,
\label{lambdaarray}
\end{eqnarray}
\end{center}
where each integration is over the Fermi surface ($FS$) either in ${\bfk}$
or ${\bfkprime}$ space.  The term $\eta({\bfk})$ is the angular part of
the superconducting order parameter.  In the above we are following the
notation adopted by Monthoux and Lonzarich \cite{monthoux_2001a} and
{\it restrict ourselves to examine majority spin triplet pairing}, using
$V_{\uparrow\uparrow}(\bfq)$.   The choice of order parameter should 
naturally
reflect the symmetry
properties of the {\ug} crystal structure.  Such considerations should
lead us to examine non-unitary states~\cite{machida_2001a, fomin_2001a},
but here for simplicity we consider as an example the states
$\Delta_{\bfk}=\Delta_{0}\sin(k_{x})$ and $\Delta_{0}\sin(k_{y})$,
anticipating that which state is favoured may change as the
majority Fermi surface is tuned through $p_x$, making the
transition from being open to being closed.  The favoured state is
determined by which $\eta({\bfk})$ gives a larger positive value of      
$\lambda_{\Delta}$.

There is a caviat at this stage.  The sharpness of the van Hove
singularities is useful in providing the magnetisation behaviour shown in
Section \ref{jumps}. However, as can be readily seen from Eq.
\ref{FayAppelV}, it can also lead to a switching of
the sign of the pair interaction from attraction to repulsion at zero    
${\bfq}$ if the density of states at the van Hove point is {\it too} high.
This is because $\chi^{(0)}_{\sigma\sigma}({\bfq}=0)=\rho(\mu_{\sigma})$.
With the presence of a strict low dimensional van Hove singularity (an 
infinity in the DOS) at zero temperature,
the quantity $\rho(\mu_{\uparrow})\rho(\mu_{\downarrow})$ will be
greater than 1 when the majority Fermi level reaches the van Hove point.

There are several possible routes to softening the van Hove singularity.
One would be to introduce a small amount of disorder into the model.
Another would be to transfer the calculation to three electronic
dimensions, rather than the current two, but that is computationally   
time-consuming. Here we take a simpler approach, calculating all
$\chi^{(0)}_{\sigma\sigma}({\bfq})$ at a small finite temperature   
$(k_{B}T=0.045)$, retaining the original zero-temperature value of $I$ for
each magnetisation examined.  This procedure maintains the order of the
magnetic transition at $I_{x}$, although it does mean that the effective
density of states,  $\chi^{(0)}_{\sigma\sigma}({\bfq=0})$ is inconsistent
with that used to find $M(I)$.  $\chi^{(0)}_{\sigma\sigma}({\bfq})$ is
calculated on a $40$ by $40$ grid for ${\bfk}$ and ${\bfkprime}$ points in
the Brillouin zone.  The agreement is close
away from the van Hove regions.

In order to have two first order magnetic phase transitions (see Fig.
\ref{CombiSuscFig}(b)), we take $N=0.77$ in what follows, and examine
majority sheet, spin triplet superconductivity. The final refinement to  
the model comes in the form of adjusting the $\bfq$-dependence of $I$ at
this stage.  Until now, we have considered a $\bfq$-independent Stoner
factor which arises from on-site repulsion of like-spin electrons.       
However, $I$ should really convey some of the physics of electron-electron
interactions at finite distances, and so, in calculations of the
superconducting potential, we convert to the following form:
\begin{equation} I\rightarrow {I \over 1+\xi q^2}
\label{StonerStruc} \end{equation} which
effectively reduces high-{$\bfq$} modes in the system, in line with the 
ferromagnetism of the real compound.  This is the simplest, first
approximation of such effects.

\begin{figure}
\includegraphics[width=\columnwidth]{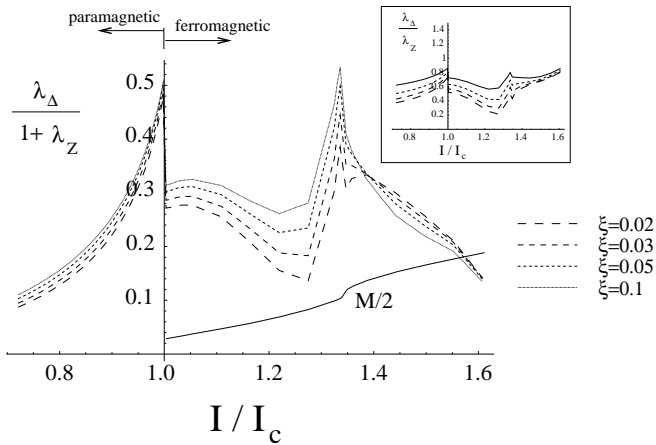}
\caption[Strength of the superconducting interaction.]{A measure of the 
strength of superconductivity: ${\lambda_{\Delta}\over 1+\lambda_{Z} }$ as 
a function of Stoner
interaction strength, $I$, normalised with respect to $I_{c}$, the
value of $I$ at the zero temperature Curie point.  $I_{x}$, the value of I
for the second jump in magnetisation, akin to the pressure identified as
$p_{x}$ in {\ug}, gives rise to the peak at $I_{x}/I_{c}\sim 1.34$.    
Also shown is the magnetisation, scaled down by a factor of two in the
dimensionless units we are using.  The
inset shows the ratio  ${\lambda_{\Delta}\over \lambda_{Z} }$ is always   
less than 1, as expected.  In both plots, we show the results for
different values of `Stoner structure factor', $\xi$ (Eq.
\ref{StonerStruc}).  The tight-binding parameters used here are again
$\beta=0.7$, $\gamma=0.03$ and $\delta=-0.03$, and the number of spins,
$N=0.77$.}
\label{LambdaDiag}
\end{figure}

\begin{figure}
\includegraphics[width=0.8\columnwidth]{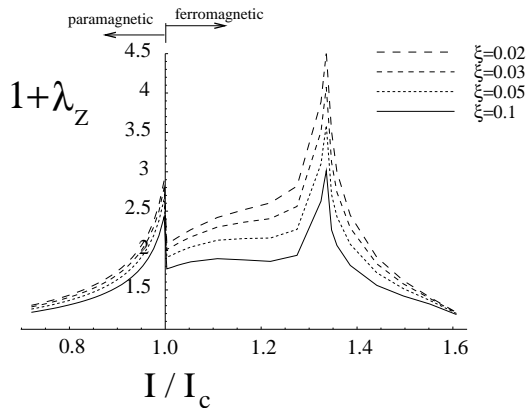}
\caption[Mass renormalisation factor.]{The mass 
renormalisation: ${1+\lambda_{Z} }$ as a function of Stoner
interaction strength, $I$, normalised with respect to $I_{c}$, the       
value of $I$ at the zero temperature Curie point.  This quantity peaks in 
the ferromagnetic state and remains high with decreasing $I$ into the
paramagnetic state.  Again, we show the results for different values of
`Stoner structure factor', $\xi$.  The tight-binding parameters used
here are again $\beta=0.7$, $\gamma=0.03$ and $\delta=-0.03$.
The number of spins, $N=0.77$.}
\label{MassFig}
\end{figure}

In Fig. \ref{LambdaDiag} we show a measure of the superconducting
interaction strength, ${\lambda_{\Delta}\over 1+\lambda_{Z} }$, for
various values of $\xi$, our Stoner `structure factor'.  The term
$1+\lambda_{Z}$ measures the mass renormalisation---this is shown
separately in Fig. \ref{MassFig}.
%%%%%%%%%%%%%%%%%%%%%%%%%%%%%%%%%%%%%%%%%%%%%%%
%%%%%%%% LAMBDA STRENGTH FIGURE        %%%%%%%%
%%%%%%%%%%%%%%%%%%%%%%%%%%%%%%%%%%%%%%%%%%%%%%%
%
%%%%%%%%%%%%%%%%%%%%%%%%%%%%%%%%%%%%%%%%%%%%%%%
%%%%%%%% LAMBDA Z (MASS) FIGURE        %%%%%%%%
%%%%%%%%%%%%%%%%%%%%%%%%%%%%%%%%%%%%%%%%%%%%%%%
Of course, ${\lambda_{\Delta}\over
1+\lambda_{Z} }$ is not the full story when one considers how the  
superconducting transition temperature $T_{SC}$
should behave as a function of $I$.  Roughly speaking, $T_{SC} \sim
\omega_{c} e^{-( { {1+\lambda_{Z} \over \lambda_{\Delta}}})}$ (see, for
example, Fay and Appel~\cite{fay_1980a}) and there is   
substantial variation of $\omega_{c}$, the paramagnon energy with $I$,
especially when soft magnetic modes are present.  The paramagnon energy
can be expanded as
\begin{equation}
\omega_{c}\sim \zeta q(\chi^{-1} + c q^{2}),
\end{equation}
where $\zeta$ and $c$ are temperature dependent parameters, equivalent to
$\gamma$ and $c$ in the work of Lonzarich~\cite{lonzarich_1997a}.  As 
developed by Brinkman and Engelsberg \cite{brinkman_1968a} and implemented 
by Fay and Appel, the
$\omega_{c}$ prefactor will substantially reduce superconductivity at the
ferromagnetic $I_{c}$ and the same argument should apply around the    
secondary transition at $I_{x}$, although in both cases this effect is
more drastic for a second order magnetic transition (where the
susceptibility diverges, i.e. $\chi^{-1}\rightarrow 0$) than for the first
order cases
here.  What remains important, therefore, is
the {\it stable region of superconductivity in the ferromagnetic state},
where ${\lambda_{\Delta}\over 1+\lambda_{Z} }$ is approximately flat and
high, especially for higher Stoner structure factor, $\xi$.

We might reasonably ask what it is that enhances superconductivity in the
region between $I_x$ and $I_c$.  In this range of $I$, the majority Fermi 
level sits between the
two van Hove peaks, and an examination of the ${\bfq}$ dependence of the
interaction potential, Eq. \ref{FayAppelV} reveals that there is a region
of high, sustained zero ${\bfq}$ pairing for 
$I_{c}<I<I_{x}$.~\cite{sandeman_2002b}  At both
$I_{c}$ and $I_{x}$, $V(\bfq)$ is broad and high around ${\bfq}=0$.  
Outwith the region $I_{c}<I<I_{x}$ the pairing potential is smaller and
more localised around zero ${\bfq}$. Furthermore, in the region between
$I_{c}$ and $I_{x}$, the mass renormalisation, represented by
$\lambda_{Z}$ is also approximately flat and high.  This mass enhancement,
shown in Fig. \ref{MassFig} compares well with the high effective 
mass plateau found in de Haas van Alphen
measurements on the ferromagnetic state between pressures $p_{c}$ and
$p_{x}$.

\begin{figure}
\includegraphics[width=\columnwidth]{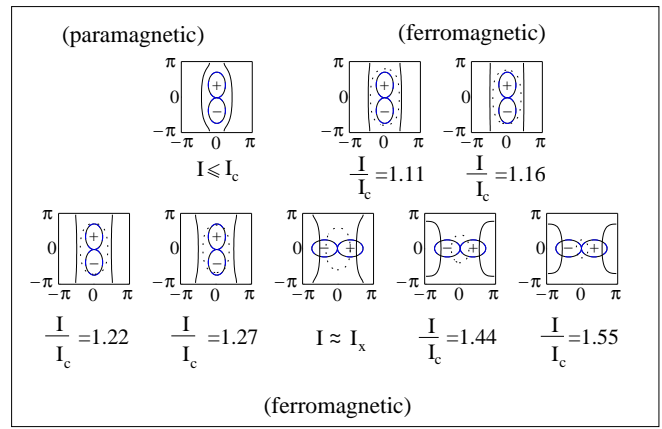}
\caption[Evolution of majority and minority Fermi surfaces as a
function of Stoner exchange.]{Evolution of majority (solid line) and
minority (dotted line) Fermi surfaces as a function of Stoner exchange. 
We see the transition in the majority sheet from an open to a closed
morphology, going through a van Hove singularity at $I=I_{x}$.  Also
shown is the order parameter used, which changes from $\sin(k_{y})$ for
$I<I_{x}$ to $\sin(k_{x})$ for $I>I_{x}$.  The tight-binding parameters
used here are again $\beta=0.7$, $\gamma=0.03$ and $\delta=-0.03$, and the
number of spins, $N=0.77$.}
\label{TopologyFig}
\end{figure}

Lastly we note that the order parameter, $\eta({\bfk})$ favoured on
tuning $I$, changes as we go through $I_x$.  This makes sense from the
point of view of the Fermi surface topology, as depicted in Figure
\ref{TopologyFig}.  $\eta({\bfk})=\sin(k_{y})$ is favoured when the    
majority spin sheet is open---for all $I<I_{x}$---and the $\sin(k_{x})$
order parameter becomes favoured for the closed surfaces for $I>I_{x}$. As
already stated, these are overly simplified gap functions, so this change 
of symmetry may well be more gradual or masked in a non-unitary
representation.

%%%%%%%%%%%%%%%%%%%%%%%%%%%%%%%%%%%%%%%%%%%%%%%
%%%%%%%% FERMI S TOPOLOGY  FIGURE      %%%%%%%%
%%%%%%%%%%%%%%%%%%%%%%%%%%%%%%%%%%%%%%%%%%%%%%%

\section{Conclusions}
In this paper we have proposed that the unusual phase diagram of {\ug} is
a result of a novel tuning of the Fermi surface topology by the
magnetisation of the ferromagnetic state.  We have constructed a model for
this which illustrates how superconductivity, the tunable magnetization
features and the quasiparticle mass are related by a twin-peak density of
states feature, consistent with experiment.  Indeed, the twin-peak DOS
feature is also apparent in the most recent LDA $+U$ calculations of the
non-magnetic state~\cite{shick_2002a}.

We have first looked for an explanation of the two transitions in the low
temperature magnetisation of {\ug}.  We have shown how a density of states
with two peaks as a function of energy will give rise generically to a 
zero temperature magnetisation which has two transitions as the ratio of 
Stoner exchange energy to bandwidth is tuned. We have demonstrated how 
this interesting form of DOS arises when the bandstructure is 
quasi-one-dimensional, as is believed to be the case for {\ug}.

We are thus able to propose a mechanism for the asymmetry of the
superconducting $T_{SC}$ with respect to the magnetic quantum phase
transition associated with the Curie temperature, shifting the focus of
our attention away from $p_{c}$ towards $p_{x}$ as the magnetic transition
point of interest with regard to superconductivity.  In our model, a large
density of states at the majority spin Fermi surface is associated with
the $T_{x}$ transition and is also driving superconductivity.  This causes
superconductivity to be favoured in the ferromagnetic state relative to
the paramagnetic state in a natural and previously overlooked way, and 
could also be a reason for the enhancement of superconductivity in the
ferromagnetic state of other materials.  We further note that, in a real
system, the presence of impurities will probably enhance further the
relative tendency to formation of the superconducting state in the
ferromagnetic phase.

There is also a straightforward explanation in the band magnetism scenario
for the observed metamagnetic behaviour and a large, constant mass
enhancement between $I_{x}$ and $I_{c}$, consistent in form with de Haas
van Alphen measurements.  Potentially we may have a reason for the
disappearance of dHvA signal between $p_{x}$ and $p_{c}$, which would here
be due to the Fermi surface becoming open.  However, we should note that
only one set of authors have observed such disappearance of dHvA
signal~\cite{settai_2002a}.

KGS wishes to thank C. Bergemann, J. Flouquet, A.D. Huxley, S.R. Julian,
A.A. Katanin, D.E. Khmelnitskii, P.B. Littlewood, A.J. Millis, P.  
Monthoux, C. Pfleiderer, S.S. Saxena, T. Schmidt, M. Sigrist and T.  
Terashima for useful discussions.  Financial support was supplied by the
EPSRC. 

\begin{thebibliography}{}
	
\bibitem{flouquet_2002a} J. Flouquet and A. Buzdin, Physics World {\bf 
15} (1):9 (2002). 
%\bibitem{shimahara_2000a} H. Shimahara and
%M. Kohmoto, Europhys. Lett. {\bf 57} 247 (2002).
\bibitem{saxena_2000a} S. S. Saxena et al., Nature (London) {\bf 406}, 587 
(2000).
\bibitem{aoki_2001a} D. Aoki et al., Nature (London) {\bf 413}, 613 
(2001).
\bibitem{pfleiderer_2001a} C. Pfleiderer et al., Nature (London) {\bf 
412}, 58
(2001).
\bibitem{lohneysen_2000a} H.V. L\"{o}hneysen 
%, C. Pfleiderer, A. Schroder and
%O. Stockert
et al., J. Phys. Soc. Jpn. {\bf 69}, 63-70 (2000).
\bibitem{mathur_1998a} N. D. Mathur et al. Nature (London) {\bf 394}, 39 
(1998).
\bibitem{huxley_2001a} A. D. Huxley et al., Phys. Rev. B {\bf 63}, 144519
(2001).
\bibitem{tateiwa_2001a} N. Tateiwa et al., J. Phys. C. {\bf
13}, L17-L23 (2001).
\bibitem{pfleiderer_2001b} C. Pfleiderer and A. D. Huxley, Phys. Rev. 
Lett. {\bf 89}, 147005 (2002).
\bibitem{oomi_1993a} G. Oomi et al., Physica B {\bf 186-188}, 758 (1993).
%\bibitem{shimizu_2001a} K. Shimizu et al., Nature (London) {\bf 412}, 316 
%(2001).
\bibitem{terashima_2001a} T. Terashima et al., Phys. Rev. Lett. {\bf 87}, 
166401 (2001).
\bibitem{fay_1980a} D. Fay and J. Appel, Phys. Rev. B {\bf 22}, 3173
(1980).
\bibitem{roussev_2001a} R. Roussev and A. J. Millis, Phys. Rev. B {\bf 
63}, 140504(R) (2001).
\bibitem{kirkpatrick_2001a} T. R. Kirkpatrick et al., Phys. Rev. Lett. 
{\bf 87}, 127003 (2001).
\bibitem{onuki_1992a} Y. Onuki et al., J. Phys. Soc. Jpn. {\bf 61}, 293
(1992).
\bibitem{shick_2001a} A. B. Shick and W. E. Pickett, Phys. Rev. Lett. {\bf
86}, 300 (2001).
\bibitem{yamagami_1993a} H. Yamagami and A. Hasegawa, Physica B {\bf
186-8}, 182 (1993).
\bibitem{tejima_2001a} S. Tejima, H. Yamagami and N. Hamada, {\it
unpublished}.
\bibitem{watanabe_2001a} S. Watanabe and K. Miyake, Physica B {\bf 312}, 
115 (2002), J. Phys. Chem. Solids {\bf 63}, 1465 (2002), {\it private 
communication} and Jpn. J. Phys. {\bf 71}, 2489 (2002). 
\bibitem{lonzarich_1985a} G. G. Lonzarich and L. Taillefer, J. Phys. C 
{\bf 18}, 4339 (1985).
\bibitem{yamada_1993a} H. Yamada, Phys. Rev. B {\bf 47}, 11211 (1993).
\bibitem{terashima_2002a} T. Terashima et al., Phys. Rev. B {\bf 65}, 
174501 (2002).
\bibitem{shimizu_1982a} M. Shimizu, J. Physique {\bf 43}, 155 (1982).
\bibitem{fazekas_1999a} P. Fazekas, {\it Lecture Notes on Electron
Correlation and Magnetism} (World Scientific, Singapore, 1999), p. 408.
\bibitem{gruner_1994a} G. Gr\"{u}ner, {\it Density Waves in Solids}
(Addison Wesley, 1994).
\bibitem{monthoux_2001a} P. Monthoux and G. G. Lonzarich, Phys. Rev. B
{\bf 63}, 054529 (2001).
\bibitem{machida_2001a} K. Machida and T. Ohmi, Phys. Rev. Lett. {\bf 86},
850 (2001).
\bibitem{fomin_2001a} I. A. Fomin, JETP Lett. {\bf 74}, 111 (2001).
\bibitem{lonzarich_1997a} G. G. Lonzarich in {\it Electron} (ed. M.
Springford), Chapter 6 (Cambridge University Press, Cambridge, 1997). 
\bibitem{brinkman_1968a} W. F. Brinkman and S. Engelsberg, Phys. Rev. {\bf 
169}, 417 (1968).
\bibitem{sandeman_2002b} K. G. Sandeman, Ph.D. thesis, University of 
Cambridge, 2002.
\bibitem{settai_2002a} R. Settai et al., J. Phys. C. {\bf 14}, L29 (2002).
\bibitem{schrieffer_1989a} J. R. Schrieffer, X. G. Wen and S. C. Zhang,
Phys. Rev. B {\bf 39}, 11663 (1989).
\bibitem{shick_2002a} A. B. Shick, {\it private communication}.
\end{thebibliography}
\end{document}